\documentclass[twocolumn,prb,reprint,superscriptaddress,floatfix,amsmath,amssymb,aps]{revtex4-2}
\pdfoutput=1

\usepackage[utf8]{inputenc}
\usepackage[caption=false,subrefformat=parens,labelformat=parens]{subfig}
\usepackage{graphicx}
\usepackage{float}
\usepackage{amsmath}
\usepackage{dcolumn}
\usepackage{bm}
\usepackage{xcolor}
\usepackage{xfrac}
\usepackage[colorlinks=true, allcolors=blue]{hyperref}
\usepackage{chemformula}
\usepackage[version=3]{mhchem}
\usepackage{siunitx}

\DeclareSIUnit\torr{torr}
\DeclareSIUnit\Oersted{Oe}

\begin{document}
\title{Absence of superconductivity in topological metal \ch{ScInAu2}}

\author{J.\ M.\ DeStefano}
\affiliation{Department of Physics, University of Florida, Gainesville FL, 32611, USA}
\author{G.\ P.\ Marciaga}
\affiliation{Department of Physics, University of Florida, Gainesville FL, 32611, USA}
\author{J.\ B.\ Flahavan}
\affiliation{Department of Physics, University of Florida, Gainesville FL, 32611, USA}
\author{U.\ S.\ Shah}
\affiliation{Department of Physics, University of Florida, Gainesville FL, 32611, USA}
\author{T.\ A.\ Elmslie}
\affiliation{Department of Physics, University of Florida, Gainesville FL, 32611, USA}
\author{M.\ W.\ Meisel}
\affiliation{Department of Physics, University of Florida, Gainesville FL, 32611, USA}
\affiliation{National High Magnetic Field Laboratory, University of Florida, Gainesville, FL 32611-8440, USA}
\author{J.\ J.\ Hamlin}
\email[Corresponding author: ]{jhamlin@ufl.edu}
\affiliation{Department of Physics, University of Florida, Gainesville FL, 32611, USA}

\date{\today}

\begin{abstract}
The Heusler compound \ch{ScInAu2} was previously reported to have a superconducting ground state with a critical temperature of \SI{3.0}{K}.
Recent high throughput calculations have also predicted that the material harbors a topologically non-trivial band structure similar to that reported for \mbox{$\beta$-\ch{PdBi2}}.
In an effort to explore the interplay between the superconducting and topological properties properties,  electrical resistance, magnetization, and x-ray diffraction measurements were performed on polycrystalline \ch{ScInAu2}.
The data reveal that high-quality polycrystalline samples lack the superconducting transition present samples that have not been annealed.
These results indicate the earlier reported superconductivity is non-intrinsic.
Several compounds in the Au-In-Sc ternary phase space (\ch{ScAu2}, \ch{ScIn3}, and \ch{Sc2InAu2}) were explored in an attempt to identify the secondary phase responsible for the non-intrinsic superconductivity.
The results suggest that elemental \ch{In} is responsible for the reported superconductivity in \ch{ScInAu2}.
\end{abstract}

\maketitle

\section{Introduction}
Many recent studies in condensed matter physics and materials science have been focused on the investigation of symmetry-protected topological states~\cite{chiu_classification_2016,po_symmetry-based_2017}.
On top of the initial efforts to identify and classify different topological states, increasing efforts have been spent on exploring the interplay between these states and other electronic and magnetic phases~\cite{wang_single_2019, ye_massive_2018}.
One such avenue of particular interest is materials systems exhibiting both non-trivial topological states and superconductivity~\cite{sato_topological_2017, qi_topological_2011}.
These compounds are candidates for being realized as true topological superconductors which are predicted to host Majorana fermions.

One such candidate, the \SI{5.4}{K} superconductor $\beta$-\ch{PdBi2}, attracted attention when it was found to have topologically non-trivial surface states~\cite{sakano_topologicalBi2Pd_2015}.
Ensuing research of the compound revealed a variety of interesting properties including complex spin textures~\cite{xu_nonhelicalPdBi2_2019} and a possible spin-triplet order parameter~\cite{iwaya_full-gapPdBi2_2017, li_observationPdBi2_2019}.
Furthermore, spectroscopic measurements on thin films of $\beta$-\ch{PdBi2} were claimed to have shown evidence of non-trivial superconductivity and Majorana fermions~\cite{lv_zeromodesPdBi2_2017}.
However, other measurements have shown that the topological surface states likely play no role in the compound's bulk superconductivity~\cite{che_andreevPdBi2_2016, biswas_notopscPdBi2_2016}.
Clearly, it would be interesting to compare these results to those for a different compound with a similar combination of superconducting and topological properties.

The search for candidate materials with certain combinations of properties has recently been facilitated by the accessibility of new databases of both experimental and computationally predicted properties.
In this case, we searched for materials that exhibited an intersection of two properties: 1. Previous experimental reports of superconductivity, and 2. Computational prediction of a topologically non-trivial band structure.
The list of experimental $T_c$ values was taken from the SuperCon database~\cite{tanifuji_supercon_database}.
Topological classification for these compounds were obtained from the the Topological Quantum Chemistry Project~\cite{bradlyn_2017, zhang_catalogue_2019, jain_commentary_2013}.
The compound \ch{ScInAu2} was among a small number of materials that indicated superconductivity at readily accessible temperatures (above $\sim \SI{2}{K}$) and a ``TI'' (topological insulator) classification.
This combination of properties lead us to investigate \ch{ScInAu2} further.
The topological classification ``topological insulator - split electronic band representation'' is the same as that for $\beta$-\ch{PdBi2}~\cite{bradlyn_2017, zhang_catalogue_2019, jain_commentary_2013}.

Given the facts above, we thus sought to characterize the potential interplay of superconductivity and topological properties in \ch{ScInAu2}.
Polycrystalline \ch{ScInAu2} was synthesized via arc-melting.
Annealing the samples yielded nearly single phase \ch{ScInAu2} that displayed no superconducting transition down to \SI{1.8}{K} via electrical resistivity and magnetization measurements.
These results are in contrast to earlier work ~\cite{matthias_obstacles_1976} which indicated superconductivity in \ch{ScInAu2} with a critical temperature of \SI{3}{\kelvin}.
Measurements reveal that only unannealed samples present the previously reported superconducting transition at \SI{3}{K}, though the shielding in the magnetic susceptibility is incomplete.
These results indicate that \ch{ScInAu2} is not superconducting down to \SI{1.8}{K} and that the previously reported critical temperature ($T_c$) of \SI{3}{K} is likely due to a secondary phase.
Based on these results several other compounds in the Au-In-Sc system were probed in search of a potential superconducting phase that could explain the partial shielding of unannealed \ch{ScInAu2} leading to the conclusion that elemental indium is responsible.

\section{Methods}
Arc-melted samples were prepared by combining the raw elements in stoichiometric ratios and melting on a water-cooled copper hearth under Ar atmosphere.
Each sample was melted multiple times, and was flipped in between each melting to ensure homogeneity throughout the boule.
Samples that did not contain In had negligible mass loss, whereas samples containing In showed mass losses around 3\%.
In order to compensate for this, extra In was added and the samples were arc-melted again until the mass of the sample indicated the correct stoichiometry had been reached. 
The samples were then annealed while wrapped in Ta foil under partial Ar atmospheres.
The crystal structures were characterized with powder x-ray diffraction (XRD) using a Siemans D500 diffractometer or a Panalytical X'Pert Pro diffractometer, and Rietveld refinements using GSAS-II ~\cite{toby_gsas-ii_2013} yielded lattice parameters consistent with those given in literature for each compound unless otherwise noted.
Electrical transport and magnetization measurements were performed in Quantum Design PPMS and MPMS systems respectively, at temperatures down to $\sim \SI{2}{K}$.

\section{Experimental Results}
\subsection{\ch{ScInAu2}}
\begin{figure}
    \centering
    \includegraphics[width=0.8\columnwidth]{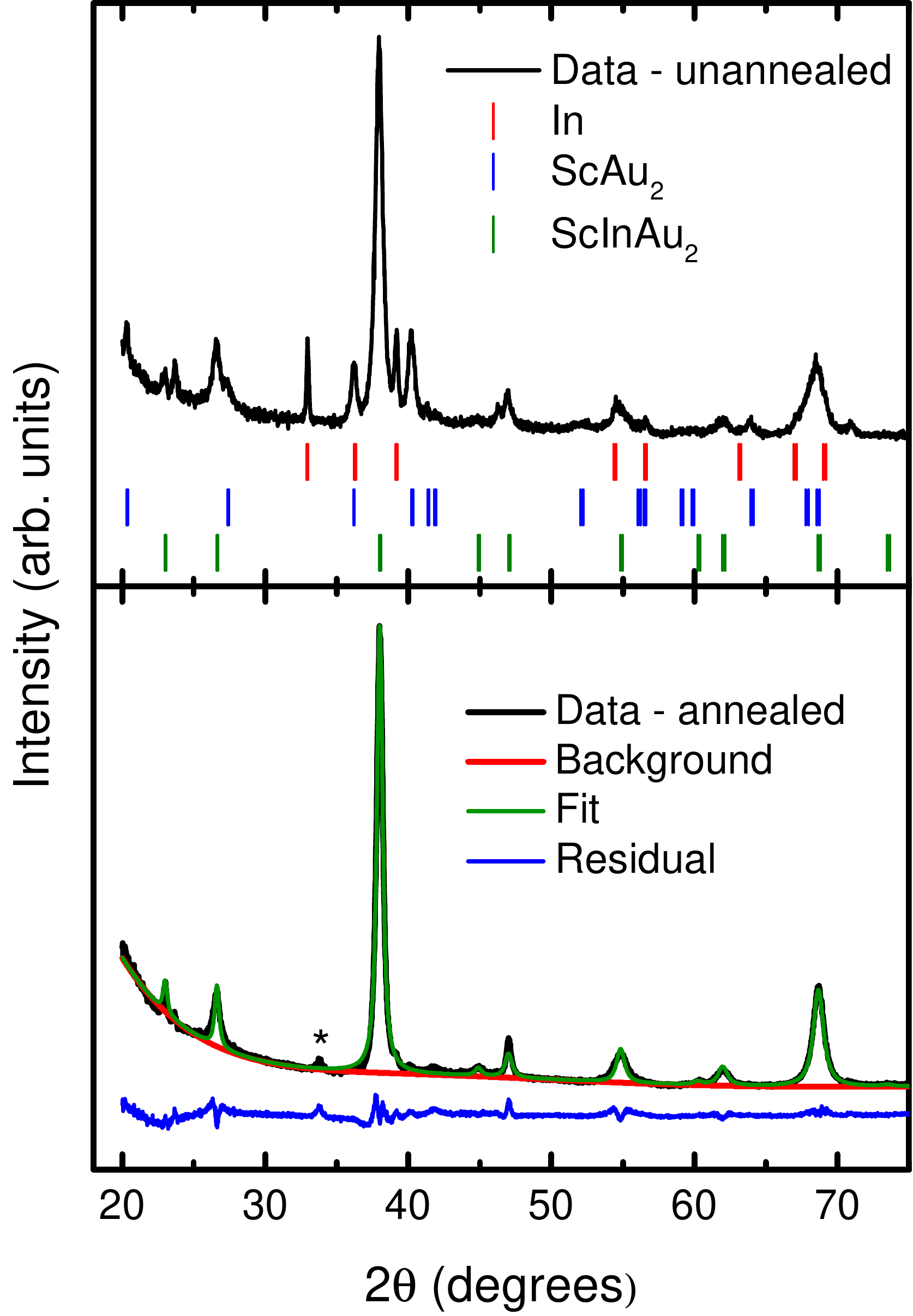}
    \caption{Top: XRD pattern of unannealed \ch{ScInAu2} with ticks indicating expected peaks of \ch{In}, \ch{ScAu2}, and \ch{ScInAu2}. Bottom: XRD pattern of annealed \ch{ScInAu2}. The small residual indicates that a nearly single-phase sample of \ch{ScInAu2} was grown. A small impurity peak is marked with an asterisk.}
    \label{fig:XRD_ScInAu2}
\end{figure}
\begin{figure}
    \centering
    \includegraphics[width=0.9\columnwidth]{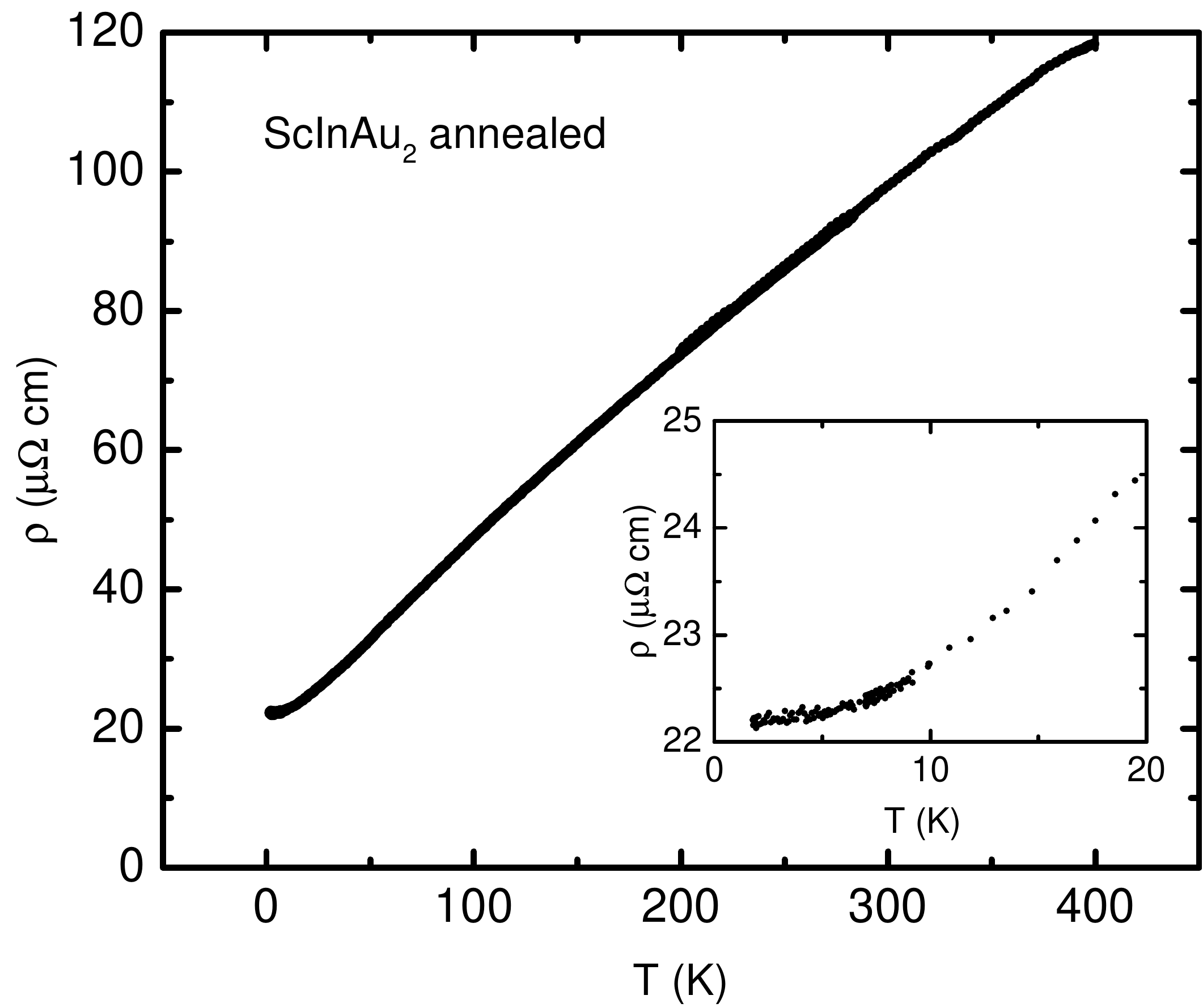}
    \caption{Resistivity versus temperature of \ch{ScInAu2} down to 1.8 K. No indication of superconductivity is observed.}
    \label{fig:resistivity_ScInAu2}
\end{figure}

Polycrystalline samples of \ch{ScInAu2} were synthesized via arc-melting.
Samples were measured both before and after annealing at \SI{700}{\celsius} for three days.
Figure~\ref{fig:XRD_ScInAu2} presents XRD data for both the annealed and unannealed samples.
While the unannealed sample shows a mixture of phases, including \ch{ScInAu2}, \ch{ScAu2}, and \ch{In}, the annealed data indicates nearly single phase \ch{ScInAu2}.
The annealed sample presents a single unidentified impurity peak near \SI{34}{\degree} (marked with an asterisk).
Electrical resistivity measurements performed on the annealed sample (Fig.~\ref{fig:resistivity_ScInAu2}) show metallic behavior from room temperature down to the base temperature of \SI{1.8}{\kelvin} with no indication of the superconductivity at \SI{3}{K} previously reported~\cite{matthias_obstacles_1976}.
It should be noted that the earlier work did not mention if the samples were subjected to any annealing process.
Therefore, we carried out additional measurements on the un-annealed multi-phase sample in order to confirm that the reported superconductivity comes from a secondary phase.

\begin{figure}
    \centering
    \includegraphics[width=\columnwidth]{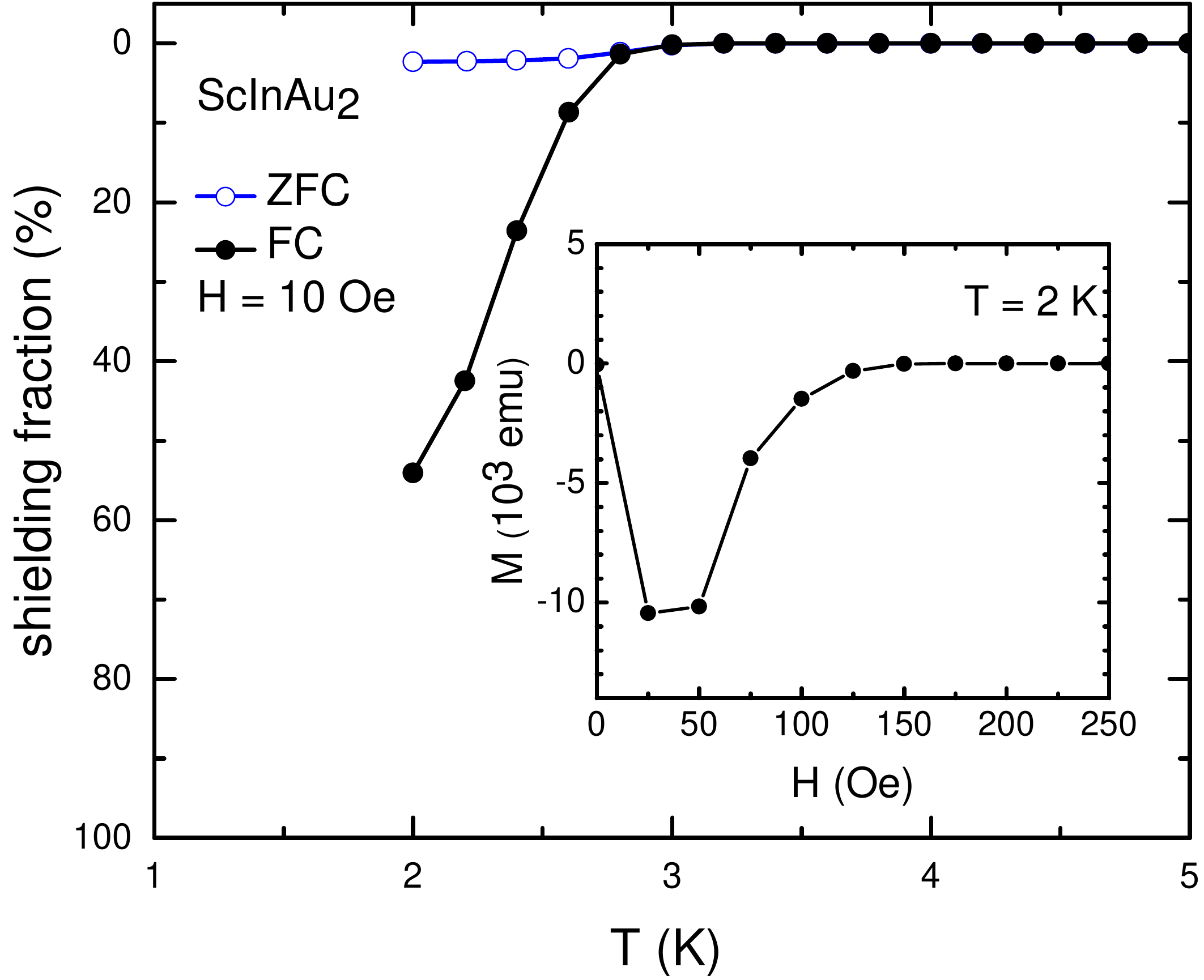}
    \caption{Shielding percentage versus temperature on unannealed \ch{ScInAu2}. The incomplete shielding suggests that an impurity phase is responsible. The inset shows the magnetization as a function of applied field.  Very small fields of order ~\SI{100}{Oe} are sufficient to suppress the superconductivity.} 
    \label{fig:magnetization_ScInAu2}
\end{figure}

Figure~\ref{fig:magnetization_ScInAu2} shows the result of magnetic susceptibility measurements on unannealed \ch{ScInAu2}.
The data show a clear drop in the susceptibility beginning slightly below \SI{3}{K}.
At the base temperature of \SI{2}{K} the transition is still incomplete but has reached a shielding fraction of more than 50\%.
In order to estimate the shielding fraction, we included the demagnetization correction of the roughly spherical sample.
The substantial shielding indicates that the secondary phase likely comprises a sizable fraction of the total sample volume.
Hence, the XRD data suggests that either In or \ch{ScAu2} is responsible.
A measurement of the magnetization vs field at \SI{2}{K} (inset of Fig.~\ref{fig:magnetization_ScInAu2}) indicates $H_{c1} ~ \sim \SI{40}{Oe}$ and complete flux expulsion by $\lesssim \SI{150}{Oe}$.
The critical field of In at \SI{2}{K} is only \SI{180}{Oe}, which is roughly consistent with our observations~\cite{crit_fields}.
The low critical field indicates that the superconducting impurity is almost certainly unreacted elemental indium ($T_c = \SI{3.4}{K}$).
Though the $T_c$ observed here is somewhat lower that that of indium ($\sim \SI{3.0}{K}$ from the onset in susceptibility), this could be caused by a combination of disorder, impurities, strain, and/or granularity.
Nonetheless, we also tested several other compounds in the Au-In-Sc system (including \ch{ScAu2}) that had not previously been measured at low temperatures in order to determine if they could instead be responsible for the superconductivity observed in the unannealed sample.

\subsection{\ch{ScAu2}}
\begin{figure}
    \centering
    \includegraphics[width=0.9\columnwidth]{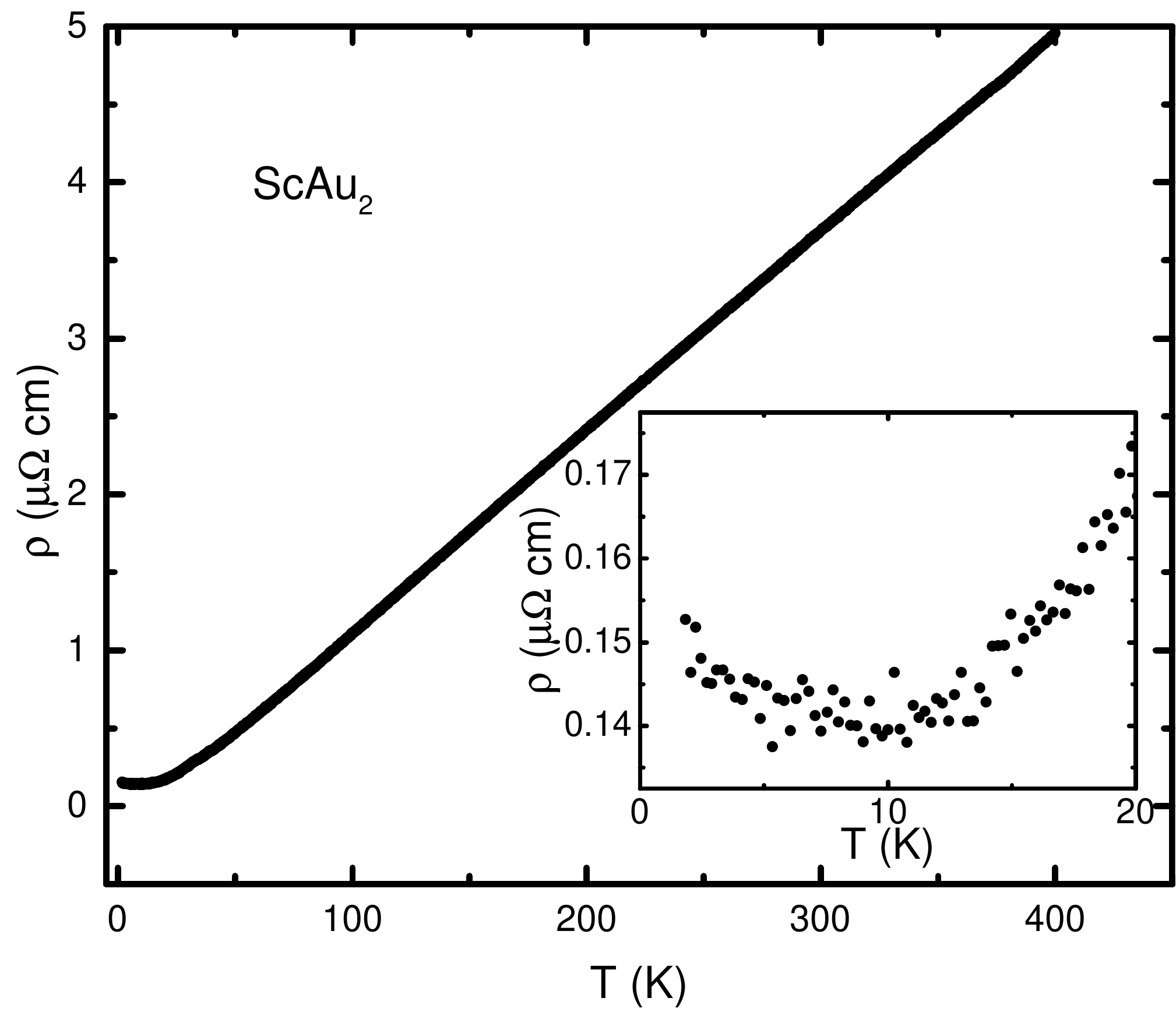}
    \caption{Electrical resistivity versus temperature for \ch{ScAu2} measured from 1.8 to \SI{400}{K}. The sample is non superconducting in this temperature range.}
    \label{fig:ScAu2 PPMS Temp Sweep}
\end{figure}

Arc melted and annealed samples of \ch{ScAu2} show diffraction patterns that matched the expected \ch{MoSi2}-type structure~\cite{Dwight1967}.
Electrical resistivity measurements present metallic behavior with a residual resistivity ratio (RRR) of $\sim 50$.
No evidence for superconductivity is detected down to \SI{1.8}{K} (see Fig.~\ref{fig:ScAu2 PPMS Temp Sweep}).
The weak upturn in resistivity below $\sim \SI{10}{K}$ could be due to a Kondo effect arising from magnetic impurities.

\subsection{\ch{ScIn3}}
Single crystals of \ch{ScIn3} were grown with the molten flux method: 80:20 atomic \% In:Sc were heated in an alumina crucible sealed in a quartz ampule under \SI{70}{\torr} \ch{Ar} gas to \SI{1000}{\celsius} and then cooled to \SI{400}{\celsius} over \SI{240}{hours}.
After holding at this temperature of \SI{8}{hours}, the ampule was centrifuged to remove the flux. This revealed small, cubic crystals, confirmed by xray diffraction to be cubic \ch{ScIn3}~\cite{Parthe1965}.

Magnetic measurements on samples yielded a diamagnetic signal with an onset of around \SI{3}{\kelvin}, but the shielding fraction of order 1\%.
Furthermore, a magnetic field of \SI{0.05}{\tesla} removed this feature.
Both of these facts indicate that the superconductivity is not intrinsic to the \ch{ScIn3} but is due to droplets of \ch{In} flux on the surfaces of the crystals.
Superconducting transitions have been observed at \SI{0.78}{\kelvin} and \SI{0.71}{\kelvin} in \ch{YIn3} and \ch{LaIn3} respectively~\cite{sharma_YIn3_LaIn3_SC}, suggesting that \ch{ScIn3} probably becomes superconducting below \SI{1}{K}.
\vspace{1em}

\subsection{\ch{Sc2InAu2}}
Samples of \ch{Sc2InAu2} were synthesized by arc melting.
The tetragonal \ch{Mo2FeB2}-type structure~\cite{HULLIGER1996160} was confirmed by x-ray diffraction, though some unidentified secondary phases were present.
Nonetheless, magnetic susceptibility measurements from 2-\SI{300}{K} presented no evidence for superconductivity or any other anomalies.

\section{Conclusions}
The previously reported superconducting behavior of \ch{ScInAu2}, a material that shares the same topological classification as $\beta$-\ch{PdBi2}, has been re-analyzed.
These measurements suggest that \ch{ScInAu2} is not intrinsically superconducting, but that unannealed samples can exhibit partial superconducting shielding in the magnetic susceptibility due to a secondary phase - most likely unreacted indium.
We also investigated the possibility that another phase is responsible for the superconductivity in unannealed samples of \ch{ScInAu2}.
Queries were performed with the Materials Platform for Data Science~\cite{andreoni_MPDS_citation} and the Superconducting Material Database~\cite{tanifuji_supercon_database} to search for compounds in the Au-In-Sc family that are reported to be superconducting.
However, no other phases with reports of $T_c \sim \SI{3}{K}$ were found.
Several compounds in this ternary phases space had not previously been characterized at low temperature, so we also screened \ch{ScAu2}, \ch{ScIn3}, and \ch{Sc2InAu2} and found that they are all essentially non-magnetic non-superconducting metals with no anomalies in the resistivity or magnetic susceptibility down to \SI{2}{K}.

With the existence of large databases of experimental and computational properties, the search for materials with certain combinations of properties is now straightforward.
In this case we identified an inaccuracy in the record - \ch{ScInAu2} is non-superconducting, though it had previously been reported to have $T_c = \SI{3}{K}$~\cite{matthias_obstacles_1976}.
However, it is clear that there are a large number of known superconducting materials with non-trivial band structures awaiting further study.

\section*{Acknowledgements}
Work on this project was supported, in part, by the National Science Foundation (NSF) via CAREER award DMR-1453752 (JJH), REU Program DMR-1852138 (GPM), DMR-1708410 (MWM), and DMR-1644779 (NHMFL), and the State of Florida.
We thank G.\ R.\ Stewart for helpful conversations.

%\section*{References cited}
\bibliography{AuInSc}

\end{document}